\documentclass[twoside]{article}
\usepackage{palatino,epsfig,latexsym}
\usepackage{hyperref}
\usepackage{amsmath}
\usepackage[auth-sc]{authblk}
\begin{document}

\title{\bf Evolution of swarming behavior\\is shaped by how predators attack}

\author[1,4,*]{Randal S.~Olson}
\author[2,4,**]{David B.~Knoester} 
\author[2,3,4.***]{Christoph Adami}

\affil[1]{Department of Computer Science \& Engineering}
\affil[2]{Department of Microbiology \& Molecular Genetics}
\affil[3]{Department of Physics and Astronomy}
\affil[4]{BEACON Center for the Study of Evolution in Action
Michigan State University, East Lansing, MI 48824, USA}
\affil[*]{rso@randaloson.com}
\affil[**]{dk@msu.edu}
\affil[***]{adami@msu.edu}

\maketitle

\begin{abstract}

Animal grouping behaviors have been widely studied due to their implications for understanding social intelligence, collective cognition, and potential applications in engineering, artificial intelligence, and robotics. An important biological aspect of these studies is discerning which selection pressures favor the evolution of grouping behavior. In the past decade, researchers have begun using evolutionary computation to study the evolutionary effects of these selection pressures in predator-prey models. The selfish herd hypothesis states that concentrated groups arise because prey selfishly attempt to place their conspecifics between themselves and the predator, thus causing an endless cycle of movement toward the center of the group. Using an evolutionary model of a predator-prey system, we show that how predators attack is critical to the evolution of the selfish herd. Following this discovery, we show that density-dependent predation provides an abstraction of Hamilton's original formulation of ``domains of danger.'' Finally, we verify that density-dependent predation provides a sufficient selective advantage for prey to evolve the selfish herd in response to predation by coevolving predators. Thus, our work corroborates Hamilton's selfish herd hypothesis in a digital evolutionary model, refines the assumptions of the selfish herd hypothesis, and generalizes the domain of danger concept to density-dependent predation.
\end{abstract}

Keywords:
group behavior,
selfish herd theory,
predator attack mode,
density-dependent predation,
predator-prey coevolution,
evolutionary algorithm,
digital evolutionary model.

\newpage

\section{Introduction}

Over the past century, researchers have devoted considerable effort into studying animal grouping behavior due to its important implications for social intelligence, collective cognition, and potential applications in engineering, artificial intelligence, and robotics~\cite{Couzin2009}. Indeed, grouping behaviors are pervasive across all forms of life. For example, European starlings ({\it Sturnus vulgaris}) are known to form murmurations of millions of birds which perform awe-inspiring displays of coordinated movement~\cite{Feare1984,Hemelrijk2011}. Western honeybees ({\it Apis mellifera}) communicate the location of food and nest sites to other bees in their group via a complex dance language~\cite{Dyer2002}. Even relatively simple bacteria exhibit grouping behavior, such as {\it Escherichia coli} forming biofilms which allow their group to survive in hostile environments~\cite{Hall-Stoodley2004}.

\emph{Swarming} is one example of grouping behavior, where animals coordinate their movement with conspecifics to maintain a cohesive group. Although swarm-like groups could arise by chance, e.g., Little Egrets ({\em Egretta garzetta}) pursuing a common resource in water pools~\cite{Kersten1991}, typically swarms are maintained via behavioral mechanisms that ensure group cohesion~\cite{Ballerini2008}. As with many traits, swarming behavior entails a variety of fitness costs, such as increased risk of predation and the requisite sharing of resources with the group~\cite{Parrish1999}. With this fact in mind, significant effort has been dedicated to understanding the compensating benefits that grouping behavior provides~\cite{Krause2002}. Many such benefits of grouping behavior have been proposed, for example, swarming may improve mating success~\cite{Yuval1993,Diabate2011}, increase foraging efficiency~\cite{Pulliam1984}, or enable the group to solve problems that would be impossible to solve individually~\cite{Couzin2009}. Furthermore, swarming behaviors are hypothesized to protect group members from predators in several ways. For example, swarming can improve group vigilance~\cite{Treherne1981,Kenward1978,Treisman1975,Pulliam1973}, reduce the chance of being encountered by predators~\cite{Treisman1975,Inman1987}, dilute an individual's risk of being attacked~\cite{Ioannou2012,Treherne1982,Foster1981,Hamilton1971}, enable an active defense against predators~\cite{Krause2002}, or reduce predator attack efficiency by confusing the predator~\cite{Ioannou2008,Jeschke2007,Krakauer1995}.

Unfortunately, many swarming animals take months or even years to produce offspring. These long generation times make it extremely difficult to experimentally determine which of the aforementioned benefits are sufficient to select for swarming behavior as an evolutionary response, let alone study the behaviors as they evolve~\cite{Jeschke2007,Beauchamp2004}. In this paper, we use a digital model of predator-prey coevolution to explore Hamilton's selfish herd hypothesis~\cite{Hamilton1971}. Briefly, the selfish herd hypothesis states that prey in groups under attack from a predator will seek to place other prey in between themselves and the predator, thus maximizing their chance of survival. As a consequence of this selfish behavior, individuals continually move toward a central point in the group, which gives rise to the appearance of a cohesive swarm. This paper expands on earlier work~\cite{Olson2013SelfishHerd} by studying the long-term evolutionary effects of differing attack modes, exploring a new attack mode that directly selects against swarming behavior, and providing an analysis of the control algorithms that evolved in the swarming prey.
\begin{figure}
\centerline{\includegraphics[width=0.5\textwidth]{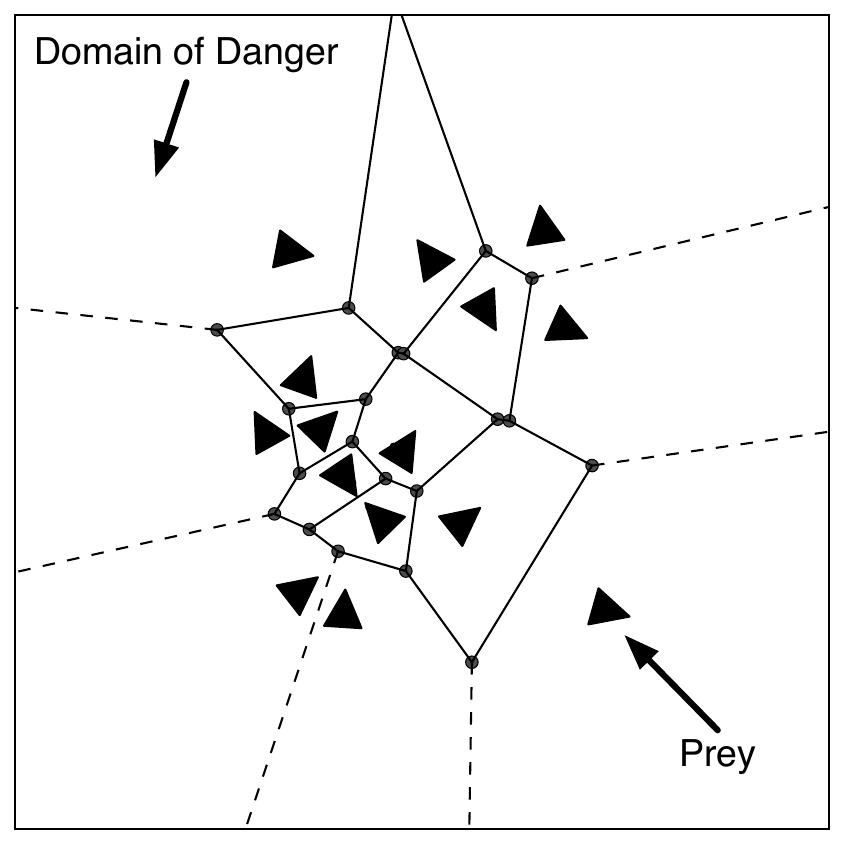}}
\caption{Example ``domains of danger'' (DODs) from Hamilton's selfish herd hypothesis. Each triangle represents a prey in the group, and the area around each triangle is its DOD. Prey on the inside of the group have smaller DODs, which means they are less likely to be targeted when a predator attacks. As a consequence, ``selfish'' prey that move inside the group to minimize their DOD will have an evolutionary advantage.}
\label{fig:domains-of-danger}
\end{figure}

\section{Related Work}

Hamilton's original formulation of the selfish herd hypothesis introduced the concept of ``domains of danger'' (DODs, Figure~\ref{fig:domains-of-danger}), which served as a method to visualize the likelihood of a prey inside a group to be attacked by a predator~\cite{Hamilton1971}. Prey on the edges of the group would have larger DODs than prey on the inside of the group; thus, prey on the edges of the group would be attacked more frequently. Moreover, Hamilton proposed that prey on the edges of the group would seek to reduce their DOD by moving inside the group, thus placing other group members between themselves and the predator. Further work has expanded on this hypothesis by adding a limited predator attack range~\cite{James2004}, investigating the effects of prey vigilance~\cite{Beauchamp2007}, considering the initial spatial positioning of prey when the group is attacked~\cite{Morrell2010}, exploring the role of prey body characteristics in shaping herd characteristics~\cite{Kunz2003,Hemelrijk2005}, and even confirming Hamilton's predictions in biological systems~\cite{Quinn2006}.

Additional studies have focused on the movement rules that prey in a selfish herd follow to minimize their DOD~\cite{Viscido2002}. This line of work began by demonstrating that the simple movement rules proposed by Hamilton reduce predation risk for prey inside the group~\cite{Morton1994}, then opened some parameters of the movement rules to evolution in an attempt to discover a more biologically plausible set of movement rules~\cite{Reluga2005,Wood2007}. Importantly, these studies demonstrated that it is possible for selfish herd behavior to evolve by natural selection on movement rules that rely on only local information for each agent, rather than global information about the entire group. This paper builds on this work by studying the effects of coevolving predators and predator attack mode (i.e., how predators select a prey in a group to attack) on the evolution of the selfish herd.

That said, there are many other potential causes of swarming behavior that we do not address in this study. For example, some studies have investigated the evolution of predator behavior in response to prey density~\cite{Tosh2011}, the role of relative predator and prey speeds on the evolution of grouping behavior~\cite{Wood2010}, elaborated upon the interaction between ecology and the evolution of grouping behavior~\cite{Spector2003,Ward2001}, and explored the role of group vigilance (i.e., the ``many eyes'' hypothesis) in the evolution of grouping behavior~\cite{Haley2014,Olson2015ManyEyes}. Two recent studies have explored the coevolution of predator and prey behavior in the presence of the predator confusion effect~\cite{Olson2013PredatorConfusion,Kunz2006PredatorConfusion}, and found that the predator confusion effect is sufficient to select for the evolution of swarming behavior in the absence of any other group benefits. Further, it has been shown that predators can adapt composite tactics to improve their efficacy against swarms of prey, which greatly reduces the defensive benefit of swarming~\cite{Demsar2015}. It is therefore necessary to keep in mind that while this study investigates the evolution of swarming behavior according to Hamilton's selfish herd hypothesis, in natural populations there are often many interacting benefits and costs of swarming behavior that must be taken into account~\cite{Krause2002}.

Of course, prior to this work there has been considerable research exploring the (co-)evolution of animal behavior in agent-based models. Craig Reynolds' work on the evolution of prey behavior in response to simulated predation was one of the earliest papers to demonstrate that predation can directly select for aggregative behavior in prey populations~\cite{Reynolds1993}. Similarly, Karl Sims' work in 1994 established a new paradigm for the coevolution of agent behavior and morphology~\cite{Sims1994}, which is still an object of intense study to this day. The work in this paper seeks to provide a stronger biological grounding to these kinds of agent-based evolution experiments, in particular by implementing specific selection pressures on the prey that are discussed in the animal behavior literature.

More broadly, in the past decade researchers have focused on the application of locally-interacting swarming agents to optimization problems, called Particle Swarm Optimization (PSO)~\cite{Poli2008}. PSO applications range from feature selection for classifiers~\cite{Xue2012}, to video processing~\cite{Vellasques2012}, to open vehicle routing~\cite{Marinakis2011}. A related technique within PSO seeks to combine PSO with coevolving ``predator'' and ``prey'' solutions to avoid local minima~\cite{Silva2002}, which has proven effective in other evolutionary computation domains as well~\cite{Angeline1993}. Researchers have even sought to harness the collective problem solving power of swarming agents to design robust autonomous robotic swarms~\cite{Sahin2004}. Thus, elaborations on the foundations of animal grouping behavior has the potential to improve our ability to solve engineering problems.

\section{Methods}

\begin{figure}
\centerline{\includegraphics[width=0.6\textwidth]{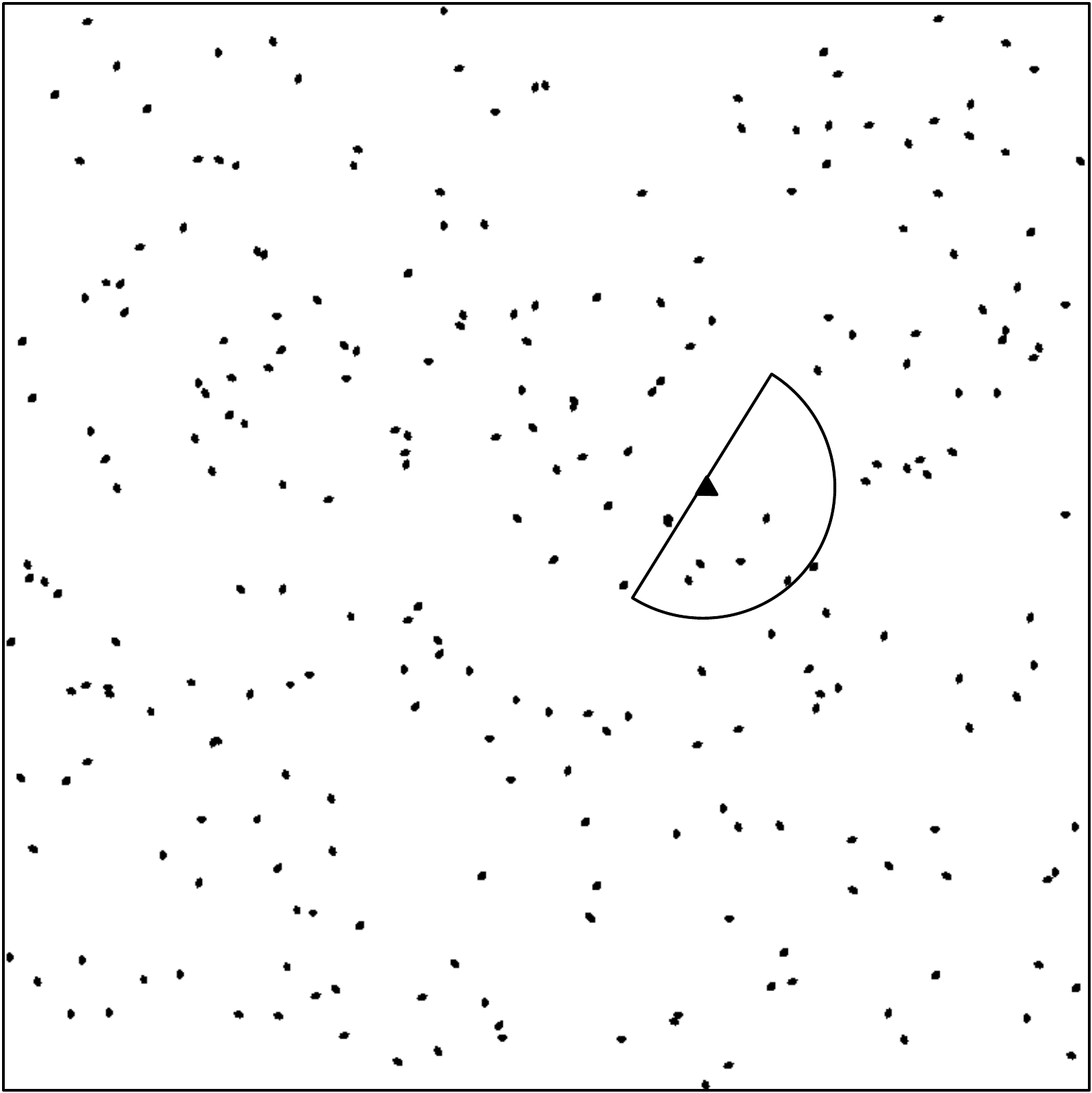}}
\caption{A depiction of the simulation environment in which the agents interact. Black dots are prey agents, the black triangle is a predator agent, and the lines around the predator agent indicate its field of view. Agents wrap around the edges of the toroidal simulation environment.}
\label{fig:sim-env}
\end{figure}

To study the evolution of the selfish herd, we developed an agent-based model in which agents interact in a continuous, toroidal virtual environment ($736\times736$ virtual meters), shown in Figure~\ref{fig:sim-env}. At the beginning of each simulation, we place 250 agents in the environment at uniformly random locations. These agents are treated as ``virtual prey.'' Each agent is controlled by a {\em Markov Network} (MN), which is a probabilistic controller that makes movement decisions based on a combination of sensory input (i.e., vision) and internal states (i.e., memory)~\cite{Edlund2011}. We evolve the agent MNs with a genetic algorithm (GA)~\cite{Eiben2003,Goldberg1989} under varying selection regimes, which will be described in more detail below.

During each simulation time step, all agents read information from their sensors and take action (i.e., move) based on their effectors. In our first set of treatments, we simulate an ideal, disembodied predator by periodically removing prey agents from the environment and marking them as consumed, e.g., when they are on the outermost edges of the group. Subsequent treatments introduce an embodied, coevolving predator agent which is controlled by its own MN. The data\footnote{Data: \href{http://dx.doi.org/10.6084/m9.figshare.663680}{http://dx.doi.org/10.6084/m9.figshare.663680}} and source code\footnote{Code: \href{https://github.com/adamilab/eos-selfish-herd}{https://github.com/adamilab/eos-selfish-herd}} from these experiments is available online for further analysis. In the remainder of this section, we describe the sensory-motor architecture of individual agents and present details related to the function and encoding of MNs.
\begin{figure}
\centerline{\includegraphics[width=0.6\textwidth]{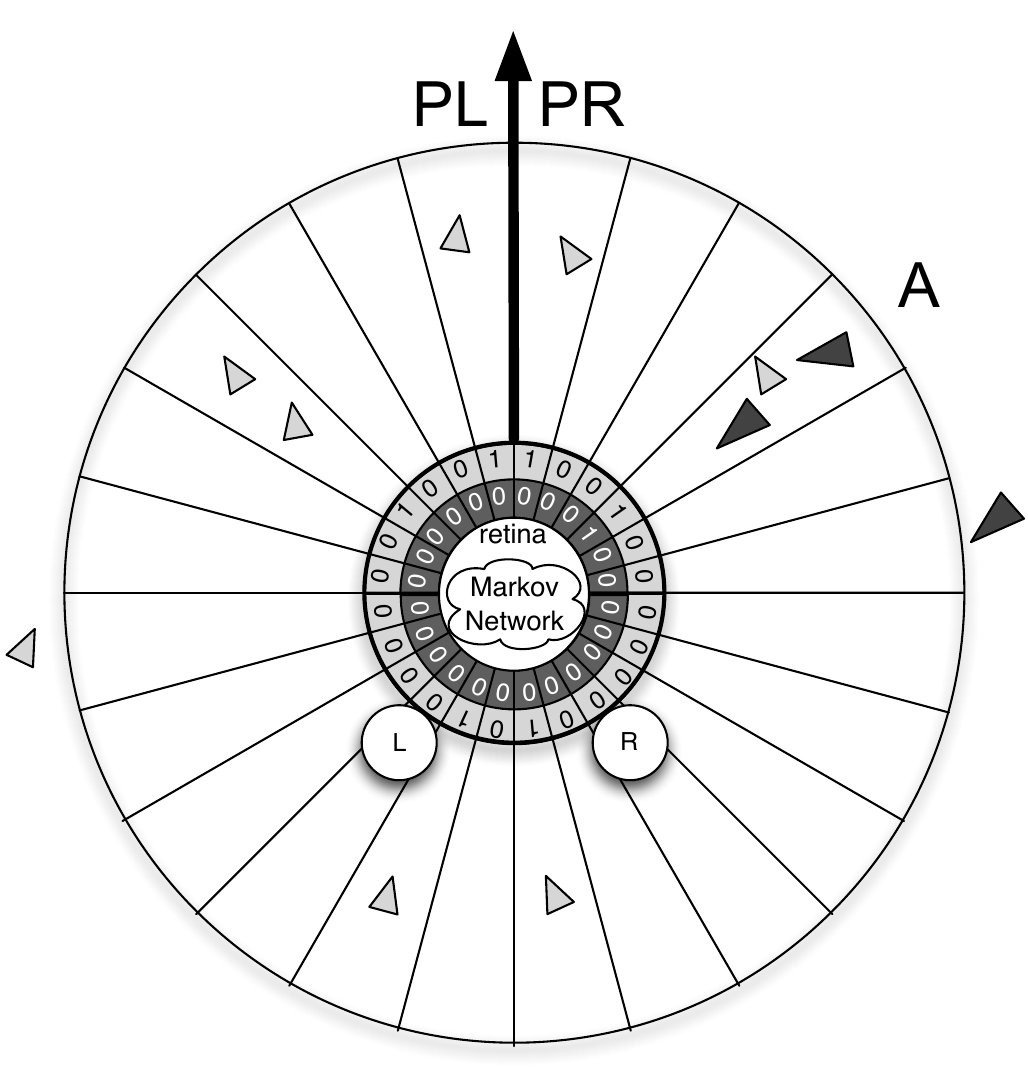}}
\caption{An illustration of the agents in the model. Light grey triangles are prey agents and the dark grey triangles are predator agents. The agents have a 360$^{\circ}$ limited-distance retina (200 virtual meters) to observe their surroundings and detect the presence of other agents. The current heading of the agent is indicated by a bold arrow. Each agent has its own Markov Network, which decides where to move next based off of a combination of sensory input and memory. The left and right actuators (labeled ``L'' and ``R'') enable the agents to move forward, left, and right in discrete steps.}
\label{fig:agent-illustration}
\end{figure}

\subsection{Agent Model}

Figure~\ref{fig:agent-illustration} depicts the sensory-motor architecture of the agents used for this study. A prey agent can sense predators and conspecifics with a limited-distance (200 virtual meters), pixelated retina covering its entire 360$^{\circ}$ visual field. Its retina is split into 24 even slices, each covering an arc of 15$^{\circ}$, which is an abstraction of the broad, coarse visual systems often observed in grouping prey~\cite{Martin1986}. Regardless of the number of agents present in a single retina slice, the prey agent only knows whether a conspecific or predator resides within that slice, but not how many. For example, in Figure~\ref{fig:agent-illustration}, the fourth retina slice to the right of the agent's heading (labeled ``A'') has both the predator and prey sensors activated because there are two predator agents and a prey agent inside that slice. Once provided with its sensory information, the prey agent chooses one of four discrete actions, as shown in Table~\ref{table:agent-action-encoding}. Prey agents turn in 8$^{\circ}$ increments and move 1 virtual meter each time step.

In our coevolution experiments, the predator agents can detect only nearby prey agents using a limited-distance (200 virtual meters), pixelated retina covering its frontal 180$^{\circ}$ that works just like the prey agent's retina (Figure~\ref{fig:agent-illustration}). Similar to the prey agents, predators make decisions about how to move next using their MN, as shown in Table~\ref{table:agent-action-encoding}, but move 3$\times$ faster than the prey agents and turn correspondingly slower (6$^{\circ}$ per simulation time step) due to their higher speed. This dramatically faster predator movement speed is meant to represent predators that perform rapid attacks on groups of prey, such as a Peregrine falcon dive bombing a swarm of Starlings. Finally, if a predator agent moves within 5 virtual meters of a prey agent that is anywhere within its retina, the predator agent makes an attack attempt on the prey agent. If the attack attempt is successful, we remove the prey agent from the simulation and mark it as consumed.

\subsection{Markov Networks}

Each agent is controlled by its own Markov Network (MN), which is a probabilistic controller that makes decisions about how the agent interacts with the environment and other agents within that environment. Since a MN is responsible for the control decisions of its agent, it can be thought of as an \emph{artificial brain} for the agent it controls. Although we specifically use MNs as the artificial brain in these experiments, other artificial brains such as Artificial Neural Networks, Genetic Programming, or many other evolvable substrates that can produce agent-based behavior based on sensory inputs could also be used in these experiments.

Every simulation time step, the MNs receive input via sensors (e.g., visual retina), perform a computation on inputs and any hidden states (i.e., memory), then place the result of the computation into hidden or output states (e.g., actuators). We note that MN states are binary and only assume a value of 0 or 1. When we evolve MNs with a GA, mutations affect (1) which states the MN pays attention to as input, (2) which states the MN outputs the result of its computation to, and (3) the internal logic that converts the input into the corresponding output.
\begin{table}
    \centering
    \caption{Possible actions encoded by the agent's output. Each output pair encodes a discrete action taken by the agent. The agent's MN changes the values stored in output states L and R to indicate the action it has decided to take in the next simulation time step.}
    \begin{tabular}{l l l}
        \hline \hline
        {\bf Output L} & {\bf Output R} & {\bf Encoded Action} \\ \hline
        0 & 0 & Move forward \\
        0 & 1 & Turn right \\
        1 & 0 & Turn left \\
        1 & 1 & Stay still \\
        \hline
    \end{tabular}
    \label{table:agent-action-encoding}
\end{table}

\subsubsection*{How Markov Networks Function}

\begin{figure}
\centerline{\includegraphics[width=0.6\textwidth]{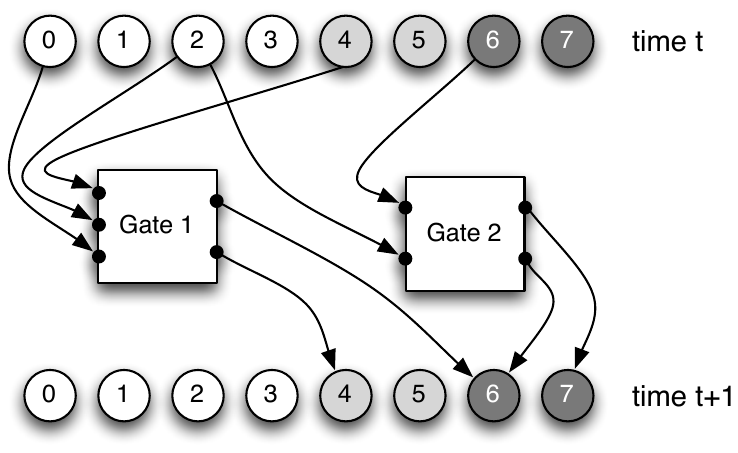}}
\caption{An example Markov Network (MN) with four input states (white circles labeled 0-3), two hidden states (light grey circles labeled 4 and 5), two output states (dark grey circles labeled 6 and 7), and two Markov Gates (MGs, white squares labeled ``Gate 1'' and ``Gate 2''). The MN receives input into the input states at time step $t$, then performs a computation with its MGs upon activation. Together, these MGs use information about the environment, information from memory, and information about the MN's previous action to decide where to move next.}
\label{fig:mnb-example}
\end{figure}

When we embed an agent into the simulation environment, we provide sensory inputs from its retina into its MN every simulation step (labeled ``retina'' and ``Markov Network'', respectively). Once we provide a MN with its inputs, we activate it and allow it to store the result of the computation into its hidden and output states for the next time step. MNs are networks of Markov Gates (MGs), which perform the computation for the MN. In Figure~\ref{fig:mnb-example}, we see two example MGs, labeled ``Gate 1'' and ``Gate 2.'' At time $t$, Gate 1 receives sensory input from states 0 and 2 and retrieves state information (i.e., memory) from state 4. At time $t+1$, Gate 1 then stores its output in hidden state 4 and output state 6. Similarly, at time $t$ Gate 2 receives sensory input from state 2 and retrieves state information in state 6, then places its output into states 6 and 7 at time step $t+1$. When MGs place their output into the same state, the outputs are combined into a single output using the OR logic function. Thus, the MN uses information from the environment and its memory to decide where to move in the next time step $t+1$.

In a MN, states are updated by MGs, which function similarly to digital logic gates, e.g., AND \& OR. A digital logic gate, such as XOR, reads two binary states as input and outputs a single binary value according to the XOR logic. Similarly, MGs output binary values based on their input, but do so with a probabilistic logic table. Table~\ref{table:mg-example} shows an example MG that could be used to control a prey agent that avoids nearby predator agents. For example, if a predator is to the right of the prey's heading (i.e., PL = 0 and PR = 1, corresponding to the second row of this table), then the outputs are move forward (MF) with a 20\% chance, turn right (TR) with a 5\% chance, turn left (TL) with a 65\% chance, and stay still (SS) with a 10\% chance. Thus, due to this probabilistic input-output mapping, the agent MNs are capable of producing stochastic agent behavior.

\begin{table}[b]
    \centering
    \caption{An example MG that could be used to control a prey agent which avoids nearby predator agents. ``PL'' and ``PR'' correspond to the predator sensors just to the left and right of the agent's heading, respectively, as shown in Figure~\ref{fig:agent-illustration}. The columns labeled P($X$) indicate the probability of the MG deciding on action $X$ given the corresponding input pair. MF = Move Forward; TR = Turn Right; TL = Turn Left; SS = Stay Still.}
    \begin{tabular}{l l|l l l l}
        \hline \hline
        {\bf PL} & {\bf PR} & {\bf P(MF)} & {\bf P(TR)} & {\bf P(TL)} & {\bf P(SS)} \\ \hline
        0 & 0 & 0.7 & 0.05 & 0.05 & 0.2 \\
        0 & 1 & 0.2 & 0.05 & 0.65 & 0.1 \\
        1 & 0 & 0.2 & 0.65 & 0.05 & 0.1 \\
        1 & 1 & 0.05 & 0.8 & 0.1 & 0.05 \\
        \hline
    \end{tabular}
    \label{table:mg-example}
\end{table}

The MGs in this model can receive input from a maximum of 4 states, and write into a maximum of 4 states, with a minimum of 1 input and 1 output state for each MG. Any state (input, output, or hidden) in the MN can be used as an input or output for a MG. MNs can be composed of any number of MGs, and the MGs are what define the internal logic of the MN. Thus, to evolve a MN, mutations change the connections between states and MGs, and modify the probabilistic logic tables that describe each MG. Mutations act directly on the genetic encoding of the MN, which is described next.

\subsubsection*{Genetic Encoding of Markov Networks}

We use a circular string of bytes as a genome, which contains all the information necessary to describe a MN. The genome is composed of {\it genes}, and each gene encodes a single MG. Therefore, a gene contains the information about which states the MG reads input from, which states the MG writes its output to, and the probability table defining the logic of the MG. The start of a gene is indicated by a {\it start codon}, which is represented by the sequence (42, 213) in the genome.
\begin{figure}
\centerline{\includegraphics[width=0.95\textwidth]{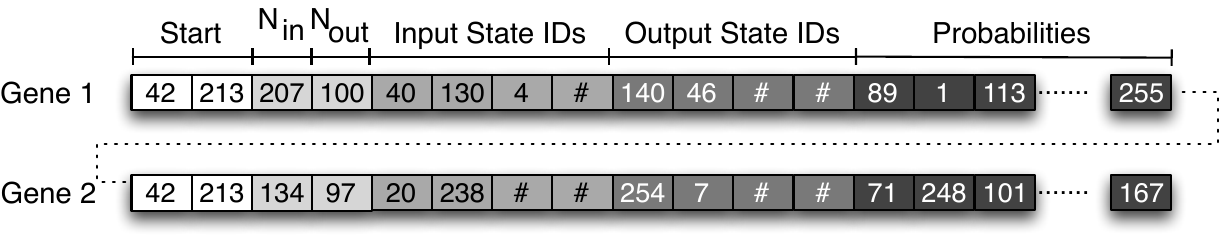}}
\caption{Example circular byte strings encoding the two Markov Gates (MGs) in Figure~\ref{fig:mnb-example}, denoted Gene 1 and Gene 2. The sequence (42, 213) represents the beginning of a new MG (white blocks). The next two bytes encode the number of input and output states used by the MG (light grey blocks), and the following eight bytes encode which states are used as input (medium grey blocks) and output (darker grey blocks). The remaining bytes in the string encode the probabilities of the MG's logic table (darkest grey blocks).}
\label{fig:mnb-encoding}
\end{figure}

Figure~\ref{fig:mnb-encoding} depicts an example genome. After the start codon, the next two bytes describe the number of inputs ($N_{\rm in}$) and outputs ($N_{\rm out}$) used in this MG, where each $N = 1 + ({\rm byte}~\mod N_{{\rm max}})$. Here, $N_{{\rm max}}$ = 4. The following $N_{\rm max}$ bytes specify which states the MG reads from by mapping to a state ID number with the equation: $({\rm byte}~\mod N_{{\rm states}})$, where $N_{{\rm states}}$ is the total number of input, output, and hidden states. Similarly, the next $N_{\rm max}$ bytes encode which states the MG writes to with the same equation as $N_{\rm in}$. If too many inputs or outputs are specified, the remaining sites in that section of the gene are ignored, designated by the \# signs. The remaining $2^{N_{\rm in} + N_{\rm out}}$ bytes of the gene define the probabilities in the logic table.

The maximum number of states allowed and which states are used as inputs and outputs are specified as constants by the user. In these experiments, we provided 64 states for the MNs to work with: 24 sensory inputs, 2 outputs for the actuators, and 38 hidden states for optional internal computations. Combined with these constants, the genome described above unambiguously defines a MN.

All evolutionary changes such as point mutations, duplications, deletions, or crossover are performed on the byte string genome, with probabilities as shown in Table~\ref{table:ga-settings}. During a point mutation, a random byte in the genome is replaced with a new byte drawn from a uniform random distribution. If a duplication event occurs, two random positions are chosen in the genome and all bytes between those points are duplicated into another part of the genome. Similarly, when a deletion event occurs, two random positions are chosen in the genome and all bytes between those points are deleted. Crossover for MNs is not implemented in this experiment to allow for a succinct reconstruction of the line of descent of the population~\cite{Lenski2003} (described more below), which is a useful tool in evolutionary studies that we harness in this paper.
\begin{table}
    \centering
    \caption{Genetic algorithm and experiment settings.}
    \begin{tabular}{l l}
        \hline \hline
        {\bf GA Parameter} & {\bf Value}\\ \hline
        Selection & Fitness proportionate\\
        Population size & 250\\
        Per-gene mutation rate & 1\%\\
        Gene duplication rate & 5\%\\
        Gene deletion rate & 2\%\\
        Crossover & None\\
        Generations & 40,000\\
        Replicates & 100 \\
        \hline
    \end{tabular}
    \label{table:ga-settings}
\end{table}

\section{Artificial Predation}

In our first set of experiments, we observe the evolution of prey behavior in response to various forms of artificial predation. This enables us to experimentally control the specific modes of predation and observe their effect on the evolution of the selfish herd. We evolve the prey genomes with a GA with the settings described in Table~\ref{table:ga-settings}. We begin the evolutionary process by seeding the prey genome pool with a set of randomly-generated ancestor genomes of length 5,000 with four random MGs. Following this, we evaluate the relative fitness of each prey genome by translating the genome into its corresponding MN, embodying each MN in a prey agent, and competing the prey agents in a simulation environment for 1,000 simulation time steps. This evaluation period is akin to the agents' lifespan, hence each agent has a potential lifespan of 1,000 time steps. We assign each prey genome an individual fitness according to how long its corresponding prey agent survived, following the equation:

\begin{equation}
    W_{{\rm prey}} = T
    \label{eq:prey-fitness-function}
\end{equation}
where $T$ is the number of time steps the prey agent survived in the simulation environment. Thus, individual prey genomes are rewarded for their agent surviving longer than other agents in the group. Once all of the prey genomes are assigned fitness values, we perform fitness-proportionate selection on the population of genomes via a Moran process~\cite{Moran1962}, increment the generation counter, and repeat the evaluation process on the new population of genomes until the final generation (40,000) is reached.

In all cases, we give the prey an initial 250 simulation time steps without predation to move around, so that prey starting on the outside of the group have the chance to move toward the center of the group if they wish to. Once the initial 250 simulation time steps elapse, we apply artificial predation every 4 simulation time steps to simulate an ideal predator attacking the group. Artificial predators succeed with their attacks every time. We limit the artificial predator attack rate to one attack attempt every 4 simulation time steps, which is called the \emph{handling time}. The handling time represents the time it takes the simulated predator to consume and digest a prey after successful prey capture, or the time it takes to refocus on another prey in the case of an unsuccessful attack attempt. We selected a handling time of 4 because it reduces the herd of prey down to 25\% of its original size by the end of the simulation, therefore applying strong selection pressure for survivorship in the herd.

For each experiment, we characterize the grouping behavior by measuring the \emph{swarm density} of the entire prey population every generation~\cite{Huepe2008}. We measure the swarm density as the mean number of prey within 30 virtual meters of each other over a lifespan of 1,000 simulation time steps, which we have experimentally shown to differentiate between swarming and non-swarming behavior in previous published experiments~\cite{Olson2013PredatorConfusion}. Qualitatively, a swarm density of $\ge15$ indicates cohesive swarming behavior, between 15 and 5 loosely grouping behavior, and $\le5$ random, non-grouping behavior. Thus, swarm density captures how cohesively the prey are swarming, or if the prey are even grouping at all.

\begin{figure}
\centerline{\includegraphics[width=0.95\textwidth]{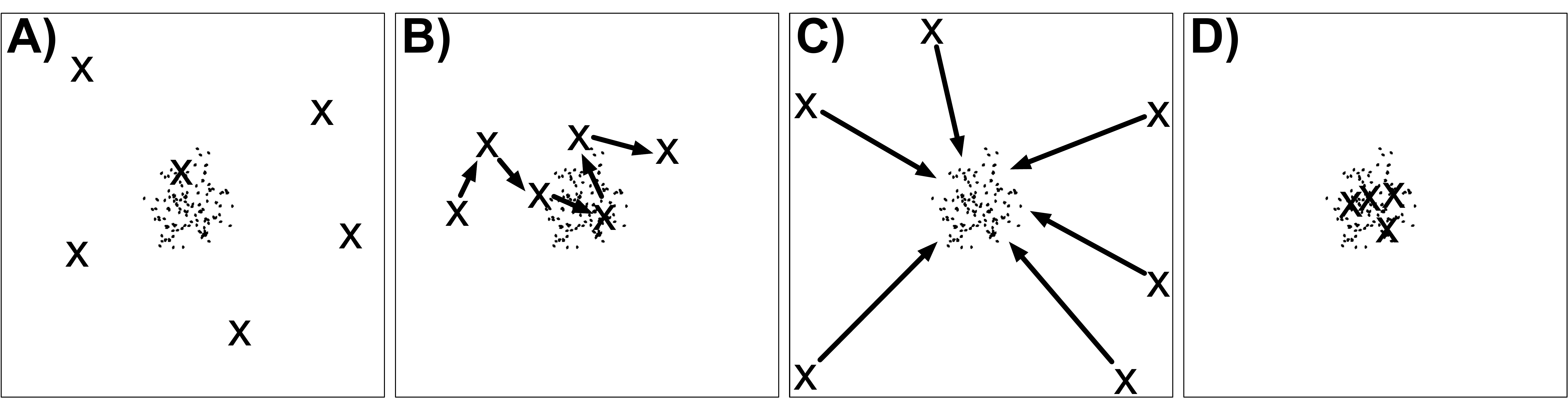}}
\caption{An illustration of the four artificial predator attack modes. A) Random attacks, B) Random walk attacks, C) Outside attacks, and D) High-density area attacks.}
\label{fig:attack-modes}
\end{figure}

In the following sections, we study the effect of four different attack modes on the evolution of swarming behavior: uncorrelated random attacks (Figure~\ref{fig:attack-modes}A), correlated random attacks (random walk attacks, Figure~\ref{fig:attack-modes}B), peripheral attacks (Figure~\ref{fig:attack-modes}C), and attacks that target the most dense area of the swarm (Figure~\ref{fig:attack-modes}D).

\subsection{Random Attacks}

Our initial study sought to verify Hamilton's selfish herd hypothesis by modeling evolving prey under attack by predators that ambush prey from a random location in the simulation environment. If the selfish herd hypothesis holds, we expect prey to minimize their ``domain of danger'' to the predators by placing as many conspecifics as possible around them~\cite{Hamilton1971}. Similar to previous models studying the selfish herd~\cite{Wood2007}, a random attack proceeds by selecting a uniformly random location inside the simulation space, then attacking the prey closest to that location, as shown in Figure~\ref{fig:attack-modes}A.

As seen in Figure~\ref{fig:sdc-artificial-selection}, swarming behavior is weakly selected for when the predators make uniformly random attacks on the prey\footnote{Video: Evolution of prey under Random Attack treatment: \href{http://dx.doi.org/10.6084/m9.figshare.658857}{http://dx.doi.org/10.6084/m9.figshare.658857}} (light grey triangles). Particularly, we found that prey took upwards of 5,000 generations to evolve cohesive swarming behavior when experiencing random attacks, compared to less than 1,000 generations with the other attack modes. However, even random attacks selected for more cohesive swarming behavior than no attacks at all, which resulted in completely dispersive behavior (Figure~\ref{fig:sdc-artificial-selection}, light grey stars).

This finding has important implications, namely that one of the original assumptions of the selfish herd hypothesis---that the predator attack mode has no impact on the evolution of swarming behavior---is not corroborated by this model. Following this discovery, we hypothesized that the \emph{directionality} of the predators' attacks play a critical role in the evolution of the selfish herd. To test this hypothesis, we next explore two different predator attack modes, each with their own distinct directionality of predation.

\subsection{Random Walk Attacks}

Our next experiment alters the mode of predation from a predator that attacks randomly selected locations to a predator that follows a random walk within the simulation environment. Shown in Figure~\ref{fig:attack-modes}B, after each attack made by this predator, it is then moved to a random location within 50 virtual meters of its previous location. This models a predator that persistently feeds on a group of prey, rather than ambushing.

\begin{figure}
\centerline{\includegraphics[width=0.95\textwidth]{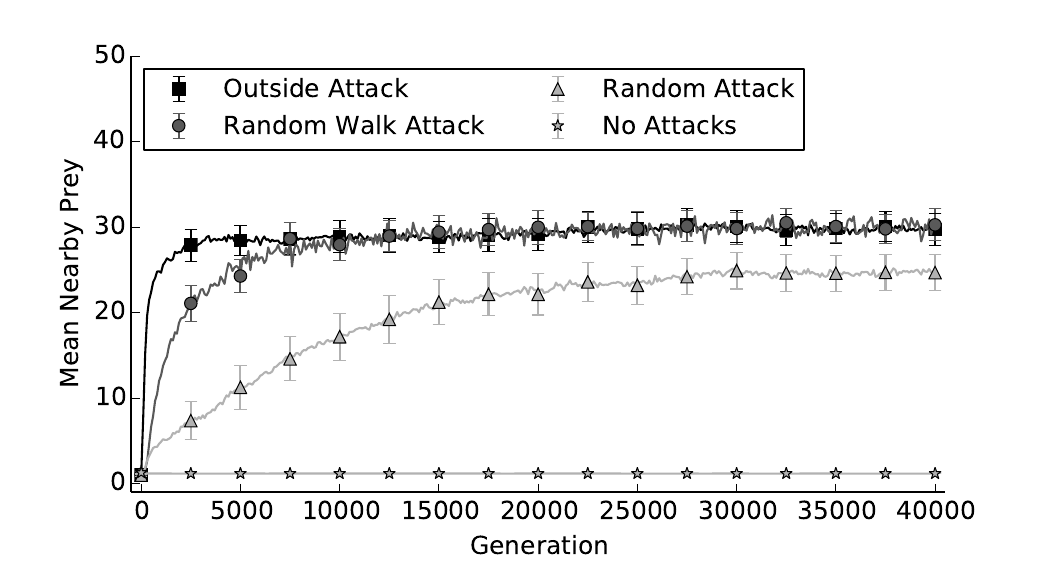}}
\caption{Mean swarm density over all replicates over evolutionary time, measured by the mean number of prey within 30 virtual meters of each other over a lifespan of 1,000 simulation time steps. Prey in groups attacked randomly (light grey triangles) took much longer to evolve cohesive swarming behavior than prey in groups attacked by a predator that follows a random walk (dark grey circles) or always from the outside of the group (black squares). When prey experience no attacks, they do not evolve swarming behavior at all (light grey stars). Error bars indicate two standard errors over 100 replicates.}
\label{fig:sdc-artificial-selection}
\end{figure}

Figure~\ref{fig:sdc-artificial-selection} shows that swarming evolved quickly when the prey were attacked by a predator following a random walk\footnote{Video: Evolution of prey under Random Walk treatment: \href{http://dx.doi.org/10.6084/m9.figshare.658856}{http://dx.doi.org/10.6084/m9.figshare.658856}} (dark grey circles). Notably, even by generation 40,000, prey experiencing random walk attacks formed significantly more cohesive swarms than prey experiencing random attacks. Thus, the random walk predator attack mode appears to capture an important aspect of predation that selects for swarming behavior.

\subsection{Outside Attacks}

In the last of our initial artificial predation experiments, we simulate a predator that always approaches from the outside of the group and attacks the prey nearest to it, as in~\cite{Viscido2001}. This predator attack mode effectively has the predators consistently attacking prey on the outer edges of the group. As shown in shown in Figure~\ref{fig:attack-modes}C, we simulate this predator attack mode by first choosing a random angle outside of the group for the predator to approach from. Once an angle is chosen, we convert the angle into a location on the edge of the visible simulation space and attack the prey nearest to that location.

As shown in Figure~\ref{fig:sdc-artificial-selection}, this form of predation has the most significant impact on the evolution of the selfish herd so far. When attacked by predators that consistently target prey on the edges of the group, prey quickly evolve cohesive swarming behavior\footnote{Video: Evolution of prey under Outside Attack treatment: \href{http://dx.doi.org/10.6084/m9.figshare.658854}{http://dx.doi.org/10.6084/m9.figshare.658854}} (black squares). Taken together, the results of these artificial predation experiments demonstrate another discovery of this work: The more predators attack prey on the outside of the group, the faster the selfish herd will evolve.

One translation of this finding is that in order for the selfish herd to evolve, prey must experience a higher predation rate on the outside of the group than in the middle of the group. While this phenomenon can be explained by each prey having a ``domain of danger'' (DOD) influenced by its relative position in the group~\cite{Hamilton1971,James2004,Morton1994}, an alternative hypothesis is that of density-dependent predation.

\subsection{Density-Dependent Predation}
\begin{figure}
\centerline{\includegraphics[width=0.95\textwidth]{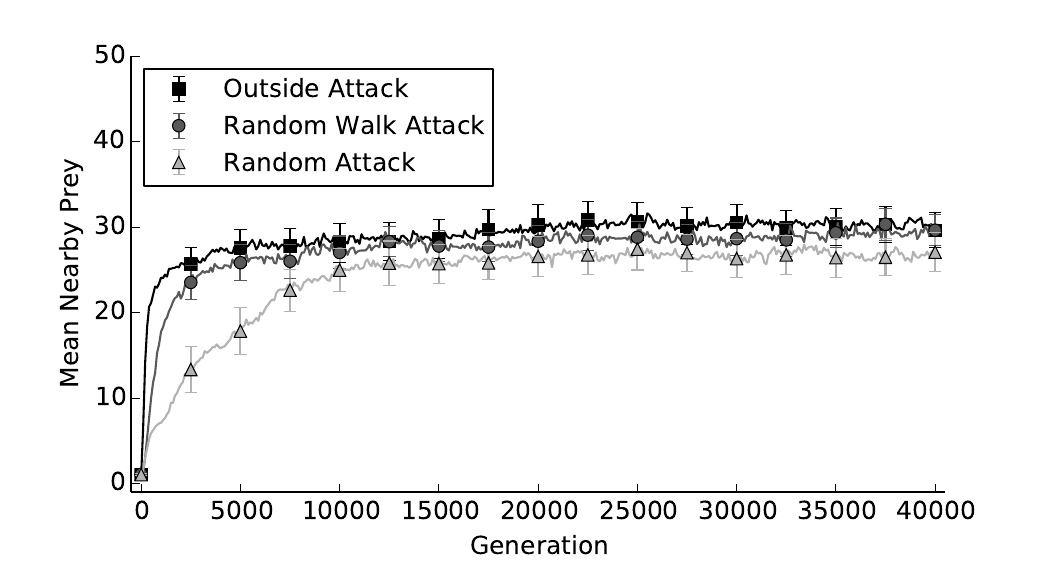}}
\caption{Mean swarm density over all replicates over evolutionary time, measured by the mean number of prey within 30 virtual meters of each other over a lifespan of 1,000 simulation time steps. Even when experiencing density-dependent predation, prey in groups attacked randomly (light grey triangles) took much longer to evolve swarming behavior than prey in groups attacked by a persistent artificial predator (dark grey circles) or always from the outside of the group (black squares). Error bars indicate two standard errors over 100 replicates.}
\label{fig:sdc-artificial-selection-dds}
\end{figure}

To study the impact of density-dependent predation on the evolution of the selfish herd, we impose a constraint on the predator which reduces its attack efficiency when it attacks areas of the group with high prey density. This reduced attack efficiency is meant to represent the increased predation rate that prey on edges of the group are expected to endure~\cite{Hamilton1971,James2004,Morton1994}, and such density-dependence can also be thought of as a proxy for group defense. We compute the predator's probability of capturing a prey during a given attack ($P_{\rm{capture}}$) with the following equation:

\begin{equation}
    P_{\rm{capture}} = \frac{1}{A_{\rm{density}}}
    \label{eq:prob-capture}
\end{equation}
where $A_{\rm{density}}$ is the number of prey within 30 virtual meters of the target prey, including the target prey itself. For example, if the predator attacks a prey with 4 other prey nearby ($A_{\rm{density}}$ = 5), it has a 20\% chance of successfully capturing the prey. As a consequence of this mechanism, the prey experience density-dependent predation.

Figure~\ref{fig:sdc-artificial-selection-dds} demonstrates the effect of density-dependent predation on the previous artificial predation experiments. Just as before, when predators did not preferentially attack prey on the outside of the group, as in the random attack experiment (light grey triangles), swarming behavior took much longer to evolve. In contrast, when the predators followed a random walk (dark grey circles) or always attacked from the outside of the group (black squares), the prey experiencing density-dependent predation again quickly evolved swarming behavior. The most noticeable effect of density-dependent predation is on the random attack treatment, where the swarm density measurement at generation 5,000 increased from 11.19$\pm$2.58 (mean $\pm$ two standard errors) to 17.61$\pm$2.72, indicating significantly stronger selection for swarming.
\begin{table}
    \centering
    \caption{High-density area attack (HDAA) experiment treatments. The values listed for each treatment are the handling times for the corresponding predator attack mode.}
    \begin{tabular}{l l l}
        \hline \hline
        {\bf HDAA?} & {\bf Outside Attack Frequency} & {\bf HDAA Frequency}\\ \hline
        No & 10 & N/A\\
        Infrequent & 10 & 250\\
        Frequent & 10 & 25\\
        \hline
    \end{tabular}
    \label{table:hdaa-treatments}
\end{table}

\subsection{High-Density Area Attacks}

Thus far, we have explored attack modes that select for the evolution of swarming behavior. It is not surprising that there are also attack modes exhibited by natural predators that must select against swarming behavior in their prey. For example, blue whales ({\it Balaenoptera musculus}) are known to dive into the densest areas in swarms of krill, consuming hundreds of thousands of krill in the middle of the swarm in a single attack~\cite{Goldbogen2011}. We call this kind of attack mode a \emph{high-density area attack}. Such an attack clearly selects against swarming behavior because it targets the prey that swarm the most. If krill swarms consistently experience these high-density area attacks, then why do they still evolve swarming behavior?

It is important to note that krill swarms are also fed on by smaller species, such as crabeater seals ({\it Lobodon carcinophagus}), that consistently attack the krill on the outside of the swarm~\cite{Mori2006}. Thus, krill swarms are experiencing two forms of attack modes simultaneously: High-density area attacks from whales and outside attacks from crabeater seals. Thus, it is possible that the selection pressure to swarm from outside attacks (Figure~\ref{fig:sdc-artificial-selection}) could outweigh the selection pressure to disperse from high-density area attacks.

Shown in Figure~\ref{fig:attack-modes}D, we model high-density area attacks as an artificial attack that always targets the prey at the most dense area of the swarm (i.e., highest $A_{\rm{density}}$). We note that this attack mode is the opposite of the density-dependent mechanism explored in the previous section, which favors predators that target prey in the \emph{least} dense area of the swarm. Once the target is selected, we execute the attack by removing the target prey and all other prey within 30 virtual meters of the target prey. Outside attacks are modeled as described above. To study the effect of high-density area attacks on the evolution of swarming behavior, we allow the prey to evolve while experiencing both attack modes simultaneously. We vary the relative handling times of both attacks (Table~\ref{table:hdaa-treatments}) to explore whether relative attack frequency could explain why some swarming animals evolved swarming behavior despite the fact that they experience high-density area attacks.

\begin{figure}
\centerline{\includegraphics[width=0.95\textwidth]{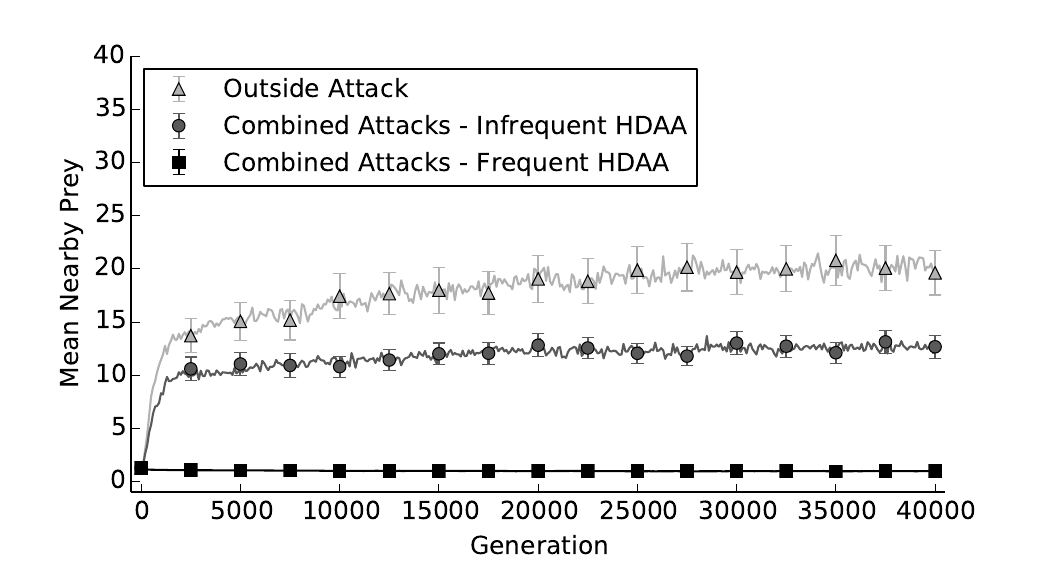}}
\caption{Mean swarm density over all replicates over evolutionary time, measured by the mean number of prey within 30 virtual meters of each other over a lifespan of 1,000 simulation time steps. Swarm density was measured from evolved populations that were not experiencing predation during measurement, eliminating any possible effects of attack modes that kill more prey faster. Prey in groups attacked only by outside attacks (light grey triangles) evolved cohesive swarming behavior. Increasing the relative frequency of high-density area attacks from infrequent (dark grey circles) to frequent (black squares) caused the prey to evolve increasingly dispersive behavior. Error bars indicate two standard errors over 100 replicates.}
\label{fig:sdc-artificial-selection-hdaa}
\end{figure}

As shown in Figure~\ref{fig:sdc-artificial-selection-hdaa}, prey experiencing only outside attacks quickly evolve cohesive swarming behavior (light grey triangles). However, when we introduce infrequent high-density area attacks (dark grey circles), the selection pressure for prey to swarm is reduced. Finally, when we introduce frequent high-density area attacks (black squares), the prey do not evolve swarming behavior at all. Thus, one possible explanation for animals evolving swarming behavior despite experiencing high-density area attacks is that the high-density area attacks are too infrequent relative to other attack types to exert a strong enough selection pressure for prey to disperse.

In summary, the artificial predation experiments provided us with two important findings regarding the evolution of the selfish herd: (1) attacks on prey on the periphery of the herd exert a strong selection pressure for prey to swarm and (2) prey in less dense areas, such as those on the outside of the herd, must experience a higher predation rate than in areas of dense prey, such as those in the center of the herd.

\section{Predator-Prey Coevolution}

Building upon the artificial predation experiments, we implemented density-dependent predation in a predator-prey coevolution experiment. Adding predators into the simulation environment enables us to observe how embodied coevolving predators affect the evolution of the selfish herd.

For this experiment, we coevolve a population of 100 predator genomes with a population of 100 prey genomes using a GA with settings described in Table~\ref{table:ga-settings}. Specifically, we evaluate each predator genome against the entire prey genome population for 2,000 simulation time steps each generation. During evaluation, we place 4 clonal predator agents inside a $512\times512$ virtual meters simulation environment with all 100 prey agents and allow the predator agents to make attack attempts on the prey agents. The prey genome population size, simulation environment area, and total number of GA generations were decreased in this experiment due to computational limitations imposed by predator-prey coevolution. We assigned the prey individual fitness values as in the previous experiments, and evaluated predator fitness according to the following equation:

\begin{equation}
    W_{{\rm predator}} = \sum_{t=1}^{t_{\rm max}} (S_0 - A_{t})
    \label{eq:predator-fitness-function}
\end{equation}
where $t$ is the current simulation time step, $t_{\rm max}$ is the total number of simulation time steps (here, $t_{\rm max}$ = 2,000), $S_0$ is the starting group size (here, $S_0$ = 100), and $A_{t}$ is the number of prey alive at update $t$. Thus, predators are selected to consume more prey faster, and prey are selected to survive longer than other prey in the group. Once all of the predator and prey genomes are assigned fitness values, we perform fitness proportionate selection on the populations via a Moran process~\cite{Moran1962}, increment the generation counter, and repeat the evaluation process on the new populations until the final generation (1,200) is reached.
\begin{figure}
\centerline{\includegraphics[width=0.95\textwidth]{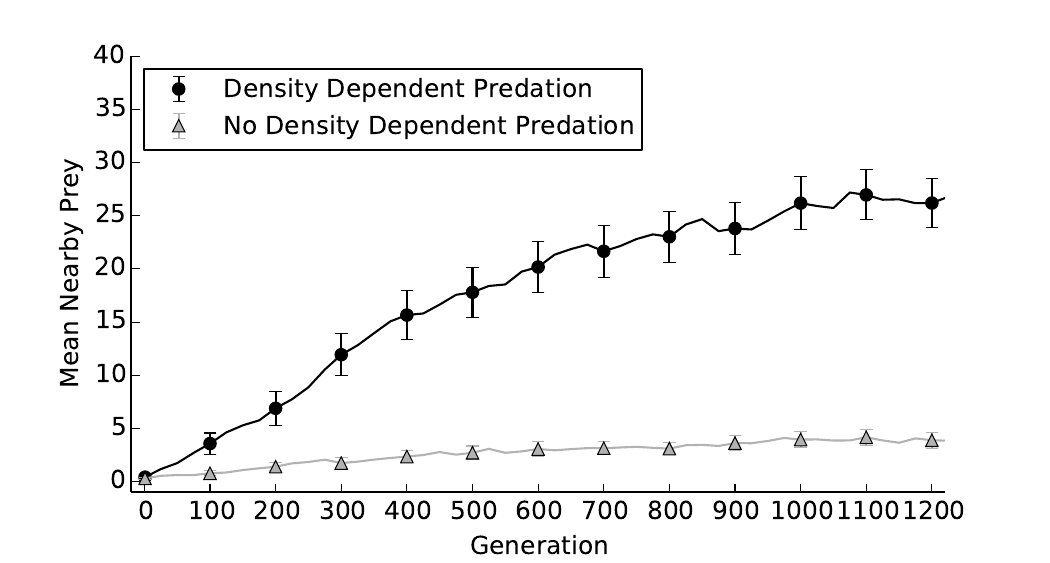}}
\caption{Mean swarm density over all replicates over evolutionary time, measured by the mean number of prey within 30 virtual meters of each other over a lifespan of 2,000 simulation time steps. Prey in groups experiencing density-dependent predation (black circles) evolved cohesive swarming behavior, whereas prey in groups not experiencing density-dependent predation (light grey triangles) evolved dispersive behavior. Error bars indicate two standard errors over 100 replicates.}
\label{fig:sdc-pred-prey-coevolution}
\end{figure}

To evaluate the coevolved predators and prey quantitatively, we obtained the line of descent (LOD) for every replicate by tracing the ancestors of the most-fit prey MN in the final population until we reached the randomly-generated ancestral MN with which the starting population was seeded (see~\cite{Lenski2003} for an introduction to the concept of a LOD in the context of digital evolution). We again characterized the prey grouping behavior by measuring the swarm density of the entire prey population every generation.

Figure~\ref{fig:sdc-pred-prey-coevolution} depicts the prey behavior measurements for the coevolution experiments with density-dependent predation\footnote{Video: Prey from predator-prey coevolution treatment: \href{http://dx.doi.org/10.6084/m9.figshare.658855}{http://dx.doi.org/10.6084/m9.figshare.658855}} (black circles; mean swarm density at generation 1,200 $\pm$ two standard errors: 26.2$\pm$2.3) and without density-dependent predation (light grey triangles; 3.9$\pm$0.8). Without density-dependent predation, the prey evolved purely dispersive behavior as a mechanism to escape the predators, even after 10,000 generations of evolution (Supplementary Figure S1). In contrast, with density-dependent predation, the prey quickly evolved cohesive swarming behavior in response to attacks from the predators within 400 generations. As expected, the coevolving predators adapted to the prey swarming behavior in the density-dependent treatments by focusing on prey on the edges of the swarm, where the density of prey is lowest. As a caveat, density-dependent predation only selects for cohesive swarming behavior when the predators are faster than the prey (Supplementary Figure S2), which corroborates earlier findings exploring the role of relative predator-prey speeds in the evolution of swarming behavior~\cite{Wood2010}.

Here we see that density-dependent predation provides a sufficient selective advantage for prey to evolve the selfish herd in response to predation by coevolving predators, despite the fact that swarming prey experience an increased attack rate from the predators due to this behavior (see ~\cite{Olson2013PredatorConfusion}, Figures S3 \& S4). Accordingly, these results uphold Hamilton's hypothesis that grouping behavior could evolve in animals purely due to selfish reasons, without the need for an explanation that involves the benefits to the whole group~\cite{Hamilton1971}. Moreover, the discoveries in this work refine the selfish herd hypothesis by clarifying the effect that different attack modes have on the evolution of the selfish herd.
\begin{figure}
\centerline{\includegraphics[width=0.55\textwidth]{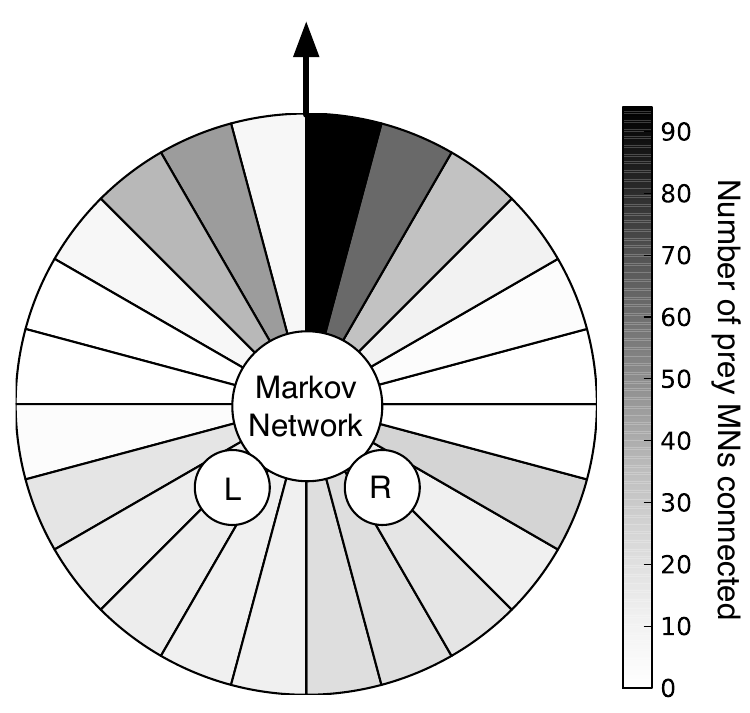}}
\caption{Number of sensory input connections from 100 evolved prey Markov Networks mapped onto a prey agent. Only causal connections from the sensory inputs to the actuators are shown. The arrow indicates the facing of the agent. The prey Markov Networks evolved a strong preference for connecting to prey sensors in front and a slight preference for sensors behind the prey agent, but tended to not connect to the sensors on the sides.}
\label{fig:prey-brain-connectivity}
\end{figure}

\section{Evolved Prey Markov Network Analysis}

Now that we have evolved emergent swarming behavior in an agent-based model under several different treatments, we can analyze the resulting Markov Networks (MNs) to gain a deeper understanding of the individual-based mechanisms underlying swarming behavior. For this analysis, we chose the most-abundant prey MN from each of the Outside Attack artificial predation experiment replicates, resulting in 100 MNs that exhibit swarming behavior. 

First, we analyze the structure of the 100 MNs by looking at the specific retina sensors that the MNs evolved to connect to. Shown in Figure~\ref{fig:prey-brain-connectivity}, the prey MNs show a strong bias for connecting to the prey-specific retina sensors in front of the prey, but not to the sides. Strangely, there appears to be a strong preference to connect to the front-right sensor but not the front-left sensor, which is an artifact of the fact that the front-right sensor is the only sensor that activates when other prey are directly in front of the prey. Additionally, some of the prey MNs show a preference for connecting to the prey-specific retina sensors behind the prey. From this analysis alone, we can deduce that the retina sensors that are most conducive to swarming behavior are in front of the prey agent.

To understand how prey make movement decisions based on their sensory inputs, we map every possible input combination in the prey's retina to the corresponding movement decision that the prey made. This mapping is accomplished by generating all $2^{24}$ possible input combinations---e.g., $000000000000000000000000$, $000000000000000000000001$, $000000000000000000000010$, etc.---and passing them into the evolved MN as a simulated sensory input. Upon activating the MN, we receive an output that corresponds to the action that the prey decided to make in respond to the simulated sensory input. Due to the stochastic nature of MNs, the prey agents do not always make the same movement decision when given the same input. Thus, we take the most-likely output from 1,000 repeats as the representative decision for a given sensory input combination. Effectively, this process produces a truth table that maps every possible sensory input to its corresponding movement decision. An example truth table can be seen in Table~\ref{tab:example-truth-table}.
\begin{table}[h]
    \centering
    \caption{An example truth table mapping every possible sensory input combination to the corresponding most-likely movement decision from the evolved prey Markov Network.}
    \begin{tabular}{l l}
        \hline \hline
        {\bf Sensory input} & {\bf Corresponding output} \\ \hline
        000000000000000000000000 & 00  \\
        000000000000000000000001 & 10  \\
        ...                      & ... \\
        111111111111111111111111 & 00  \\
        \hline
    \end{tabular}
    \label{tab:example-truth-table}
\end{table}

Once we have the truth table of all $2^{24}$ input-output mappings, we pass the truth table to the logic minimization software \emph{espresso}~\cite{Rudell1986}, which eliminates the inputs that have no effect on the outputs and provides the minimal representative logic of the truth table. This process results in a truth table that is reduced enough to make the evolved prey behavior comprehensible by humans. An example output minimal logic table can be seen in Table~\ref{tab:example-minimal-logic-table}.
\begin{table}[h]
    \centering
    \caption{An example minimal logic table resulting from a Markov Network. Input 12 corresponds to a frontal sensor, whereas Input 19 corresponds to a back-right sensor, where $1$ indicates that the sensor detects a prey agent and $0$ means no prey are in that sensor. The output $01$ translates into the agent turning right and the output $00$ translates into the agent moving forward. Thus, this example agent moves forward if it sees anything in its frontal sensor. Otherwise, the agent turns right if it sees another prey in its back-right sensor or if it sees nothing at all.}
    \begin{tabular}{l l l}
        \hline \hline
        {\bf Input 12} & {\bf Input 19} & {\bf Corresponding output} \\ \hline
        0 & 0 & 01 \\
        1 & 0 & 00 \\
        0 & 1 & 01 \\
        1 & 1 & 00 \\
        \hline
    \end{tabular}
    \label{tab:example-minimal-logic-table}
\end{table}

Surprisingly, the individual-based mechanisms underlying the emergent swarming behavior are remarkably simple. Most of the prey MNs evolved to make their movement decisions based off of only one prey sensor in front of the prey agent. If the prey sensor does not detect another prey agent, the agent repeatedly turns in one direction until it detects another prey agent in that sensor. Once the agent detects another prey agent in the sensor, it moves forward until the agent is no longer visible. This mechanism alone proved sufficient to produce cohesive swarming behavior in the majority of our experiments. Interestingly, this discovery corroborates the findings in earlier studies suggesting that complex swarming behavior can emerge from simple movement rules when applied over a population of locally-interacting agents~\cite{Olson2013PredatorConfusion,Vicsek1995,Reynolds1987}.

In a small subset of the evolved prey MNs, we observe MNs that occasionally connect to one of the prey sensors behind them. These MNs watch for a prey agent to appear in a single prey sensor behind the agent and turn repeatedly in one direction until a prey agent is no longer visible in that sensor. Once a prey agent is no longer visible in the back sensor, the MN moves forward or turns depending on the state of the frontal sensor. We note that this mechanism \emph{only} evolved in prey MNs that already exhibited swarming behavior using one of the frontal sensors, which suggests that this mechanism does not play a major role in swarming behavior. Instead, this mechanism seems to cause the prey agent to turn toward the center of the swarm instead of swarming in a circle with the rest of the prey agents. This mechanism can be thought of as a ``selfish herd'' mechanism that attempts to selfishly move the agent toward the center of the swarm to avoid predation.

\section{Conclusions and Future Work}

The contributions of this work are as follows. First, we demonstrate Hamilton's selfish herd hypothesis in a digital evolutionary model and highlight that it is the attack mode of the predator that critically determines the evolvability of swarming behavior. Second, we show that density-dependent predation is sufficient for the selfish herd to evolve as long as the predators cannot consistently attack prey in the center of the group. Finally, we show that density-dependent predation is sufficient to evolve grouping behavior in prey as a response to predation by coevolving predators. Consequently, future work exploring the evolution of the selfish herd in animals should not only consider the behavior of the prey in the group, but the attack mode of the predators as well. Following these experiments, we analyze the evolved control algorithms of the swarming prey and identify simple, biologically-plausible agent-based algorithms that produce emergent swarming behavior, including a mechanism that produces ``selfish'' behavior that drives the prey toward the center of the swarm.

Of course, the evolved prey behavior shown in the videos accompanying this paper may not closely resemble the anti-predator behavior of many species of group-living prey that we observe in nature. We provide these videos to demonstrate that grouping-like behavior has indeed evolved---and to confirm that the swarm density count metric accurately captures when the prey evolve grouping-like behavior---but we do not seek to claim that we have evolved a particular behavioral phenotype that would match the grouping behavior we observe in nature. Presumably such grouping behavior has been selected for by a variety of environmental factors that are not completely captured in this model, which would make a fascinating venue of research in the future.

While this work shows one method by which the the evolution of grouping behavior can be studied, there remain many different hypotheses explaining the evolution of grouping behavior~\cite{Krause2002}. Our future work in this area will focus on directly exploring these hypotheses in similar digital evolutionary models, which has been detailed in~\cite{Olson2015Thesis}.

\section{Acknowledgments}

We thank the three anonymous reviewers for their many insightful comments and suggestions. This research has been supported in part by the National Science Foundation (NSF) BEACON Center under Cooperative Agreement DBI-0939454, by NSF grant OCI-1122617, and by Michigan State University through computational resources provided by the Institute for Cyber-Enabled Research. Any opinions, findings, and conclusions or recommendations expressed in this material are those of the author(s) and do not necessarily reflect the views of the NSF.


\clearpage
\makeatletter 
\renewcommand{\thefigure}{S\@arabic\c@figure} 
\renewcommand{\thetable}{S\@arabic\c@table}
\setcounter{figure}{0}

\noindent{\Large Supplementary Text}

\noindent {\large \bf Olson et al.: Evolution of swarming behavior is shaped by how predators attack.}

\medskip

\subsection*{Predator-prey coevolution experiments for 10,000 generations}

\begin{figure}[h]
\centerline{\includegraphics[width=0.95\textwidth]{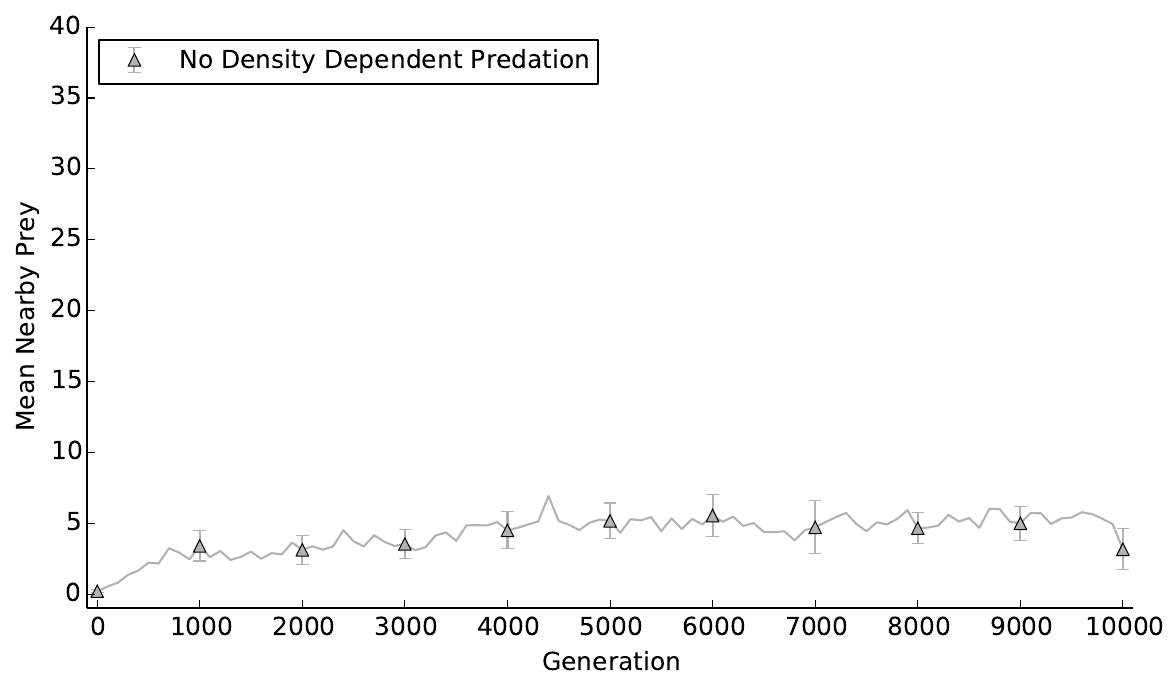}}
\caption{Mean swarm density over all replicates over evolutionary time, measured by the mean number of prey within 30 virtual meters of each other over a lifespan of 2,000 simulation time steps. Prey under the ``No Density Dependent Predation'' treatment never evolves swarming behavior even after 10,000 generations of evolution. Error bars indicate two standard errors over 30 replicates.}
\label{fig:sdc-coev-10kgen}
\end{figure}

To ensure that the ``No Density Dependent Predation'' treatment never evolves swarming behavior, we ran this treatment out for a full 10,000 generations, or over 8x the original number of generations. Figure~\ref{fig:sdc-coev-10kgen} depicts the result of these experiments, where the prey populations evolve dispersive behavior by generation 1,000 and maintain the same behavior over the whole evolutionary time span.

\newpage

\subsection*{Effect of relative predator-prey speed in predator-prey coevolution experiments}

\begin{figure}[h]
\centerline{\includegraphics[width=0.95\textwidth]{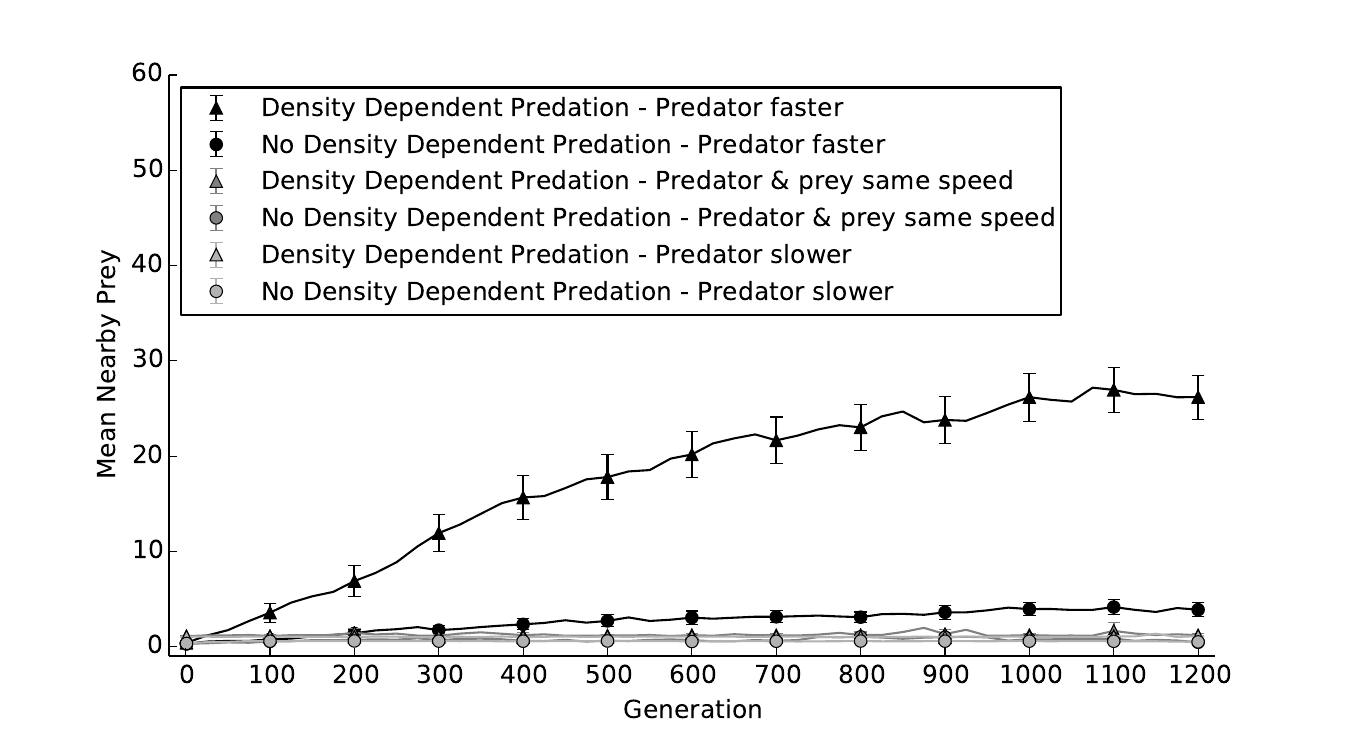}}
\caption{Mean swarm density over all replicates over evolutionary time, measured by the mean number of prey within 30 virtual meters of each other over a lifespan of 2,000 simulation time steps. Prey only evolve swarming behavior when the predator is faster and they are experiencing density-dependent predation. Error bars indicate two standard errors over 30 replicates.}
\label{fig:sdc-coev-speed-comp}
\end{figure}

To explore the role of relative predator and prey speed in our evolutionary digital model, we reran the predator-prey coevolution experiments with and without density-dependent predation and with varying relative predator and prey speeds. In the ``Predator faster'' treatment, the predators move 3x faster than the prey, whereas in the ``Predator slower'' treatment, the predators move 0.5x the speed of the prey. As shown in Figure~\ref{fig:sdc-coev-speed-comp}, regardless of whether density-dependent predation is in effect, swarming behavior only evolves in the prey when the predators are faster than the prey. In the treatments where the predator is the same speed or slower, the prey are simply able to outrun the predators, and therefore do not need to evolve swarming behavior as a defensive response to predation.

\
\end{document}